\documentclass[superscriptaddress,numbers,
 amsmath,amssymb,
 aps,
prb,
twocolumn,
]{revtex4-2}

\usepackage{graphicx}
\usepackage{multirow}
\usepackage{svg}
\usepackage{booktabs}
\usepackage[colorlinks=true, linkcolor=blue, citecolor=blue]{hyperref}%
\usepackage{natbib}
\begin{document}
\setcitestyle{sub} 
\title{Properties of 2D Electron or Hole Gases at Tailored s-Si/SiGe Interfaces: A First-Principles Investigation}

\author{Garima Ahuja} 
\affiliation{Theoretical Sciences Unit and School of Advanced Materials, Jawaharlal Nehru Centre for Advanced Scientific Research, Jakkur, Bengaluru 560064, India}

\author{Farha S}
\affiliation{School of Physics, Indian Institute of Science Education and Research Thiruvananthapuram, Maruthamala PO, Vithura, Kerala, 695551, India}
\affiliation{Center for High Performance Computing, Indian Institute of Science Education and Research Thiruvananthapuram, Maruthamala PO, Vithura, Kerala, 695551, India}

\author{Tanuja Shridhar Joshi} 
\affiliation{Theoretical Sciences Unit and School of Advanced Materials, Jawaharlal Nehru Centre for Advanced Scientific Research, Jakkur, Bengaluru 560064, India}

\author{Anil Shaji}
\affiliation{School of Physics, Indian Institute of Science Education and Research Thiruvananthapuram, Maruthamala PO, Vithura, Kerala, 695551, India}
\affiliation{Center for High Performance Computing, Indian Institute of Science Education and Research Thiruvananthapuram, Maruthamala PO, Vithura, Kerala, 695551, India}

\author{Shobhana Narasimhan}
\affiliation{Theoretical Sciences Unit and School of Advanced Materials, Jawaharlal Nehru Centre for Advanced Scientific Research, Jakkur, Bengaluru 560064, India}
\affiliation {International Centre for Theoretical Sciences, Shivakote, Hesaraghatta, Bengaluru 560089, India}

\begin{abstract}

We have performed first-principles hybrid density functional theory calculations to study the formation and properties of two-dimensional electron or hole gases (2DEG or 2DHG) at s-Si/SiGe interfaces. For small Ge concentrations $x < 0.25$, we find a novel type of band alignment with no offset in the conduction bands, implying that a 2DEG cannot be formed, though a 2DHG can. In contrast, for $x > 0.25$ the band alignment suggests that either a 2DEG or 2DHG can be formed. The electronic band structure features two nearly degenerate 2DEG states at the bottom of the conduction bands, and two 2DHG states at the top of the valence band. These states can be accessed by appropriate doping and gating. Charge density plots of these states show that they feature carriers confined to the near vicinity (2--3 atomic layers) of the interface. Calculated effective masses are anisotropic, being markedly so for the 2DHG states, and in excellent agreement with experiment. This property can be exploited to create a 1D carrier gas. Our results are especially important for  s-Si/SiGe-based  semiconducting spin qubits for quantum computing applications.
\newline
\newline
Correspondence: Shobhana Narasimhan (shobhana@jncasr.ac.in)
\newline
\newline
Submitted to Advanced Functional Materials. (Wiley)

\end{abstract}

\maketitle
\date{today}

\section{Introduction}

Semiconductor heterostructures allow one to engineer materials with tailored properties that cannot be achieved with a single semiconductor alone. By combining two or more distinct semiconductors  one can construct devices such as transistors, lasers and solar cells.
Among the various interesting phenomena that can occur at or near the heterostructure interface is the formation
of a two-dimensional electron gas (2DEG) or two-dimensional hole gas (2DHG). In these, electrons or holes are confined to a region near the interface, with a very small width ($\sim$ nm) perpendicular to the plane of the interface. The 2DEG or 2DHG is fascinating from a fundamental point of view \cite{Yu-Cardona}, possibly exhibiting exotic phenomena such as the Integer and Fractional Quantum Hall effects, Spin Hall effect, Wigner crystallization, and non-Abelian anyons \cite{klitzing,qhe_Lai2004,she_sinova2015,anyons_stern2008}. It can also be exploited for various applications, and plays an important role in the operation of electronic devices such as n- and p-channel MOSFETs \cite{MOSFET}. 

Though the possibility of engineering a 2DEG or 2DHG at semiconductor hetero-interfaces has been known for four decades now, this field is currently receiving renewed attention. This is largely because of the realization that a 2DEG or 2DHG at a semiconductor interface can be tailored to construct semiconducting spin qubits used for quantum computing. By appropriate gating, the 2DEG or 2DHG is broken up into quantum dots. In the conventional semiconducting spin qubits, each quantum dot is then further tuned to contain a single unpaired electron or hole, which constitutes the qubit \cite{Loss-DiVincenzo, scapucci}. Alternatively, in the recently developed exchange-only qubits the joint spin state of single
electrons in three neighboring quantum dots constitutes the qubit \cite{exc_qubit,exc_qubit2,exc_qubit3}. In this context, the questions of current interest are first, whether one wants to start working with a 2DEG or a 2DHG, and then which materials the heterostructure should therefore ideally be composed of. Initial efforts to make semiconducting spin qubits used a 2DEG. Now however, both 2DEG-based and 2DHG-based devices are being explored \cite{hole-qubits}. Advantages of the latter are that stronger spin-orbit interaction in the relevant systems makes the qubits easier to control, and they also lack the problem related to valley splitting in the conduction bands \cite{hole-qubits, hole-qubits-2}. However, strong spin-orbit interactions also lead to increased charge noise and thus greater decoherence \cite{coherence_chirolli01052008}.

There are two main mechanisms that can be responsible for the formation of a 2DEG/2DHG at semiconductor interfaces. For non-polar semiconductors, if there is an appropriate  band alignment between the valence band (VB) and conduction band (CB) edges of the two semiconductors, then, in the presence of modulation doping \cite{dingle_mod_doping}, a 2DEG or 2DHG can be formed \cite{abstreiter-2deg, people-2dhg, stormer,stormer_2deg}. In contrast, in polar semiconductors, the 2DEG or 2DHG results from a polarization discontinuity at
the interface that induces an electrostatic instability that is
relieved by electronic reconstruction and carrier accumulation near the interface \cite{ambacker1999}. In this paper, we restrict ourselves to the case of non-polar semiconductors.

The 2DEG and 2DHG were first experimentally engineered in the GaAs/AlGaAs system \cite{stormer,stormer_2deg}, and early attempts to build semiconducting spin qubits continued to work with these materials. This was primarily because of the availability of high-purity samples with low disorder and low scattering. The electronic structure of the system also presented advantages, such as the low electronic effective mass in GaAs. However, this system also has disadvantages. GaAs lacks a nuclear-spin-free isotope, and the strong hyperfine interaction leads to noise and causes decoherence of the qubit \cite{scapucci}. It is also not CMOS-compatible, making large-scale integration difficult.

For these reasons, and bolstered by the recent availability of isotopically pure Si \cite{edllbauer2025}, several present day attempts to build semiconducting spin qubits have used heterostructures built out of SiGe alloys and strained-Si (s-Si) -- we note that this should be distinguished from strained-SiGe/Si, which is also a system of interest \cite{guldner,Yang, paul}.
Unlike GaAs, Si can be processed to yield a nuclear-spin-free isotope. Si- and Ge-based qubits are also compatible with the well developed semiconductor manufacturing technology; this can be leveraged
to achieve scalable quantum computing architectures. High fidelity in Si based qubits has been observed, making these highly suitable for quantum computation applications \cite{fidel_si_2}.

Si/SiGe heterostructures can host a 2DEG (frequently reported) or 2DHG (rarely reported) at the interface \cite{abstreiter-2deg, people-2dhg,2deg_30_sb_doped,whall,vogl_, eff_mass_measurement,eff_mass_song,fidorra_2deg}. We note that in some earlier reports, it is not clear whether the strain is in Si or in SiGe.
Given the decades of earlier work on this system, it is at first surprising to realize that there has been a dearth of first principles computational investigations of this system. Crucial parameters, such as the Ge concentration $x$ in the Si$_{1-x}$Ge$_x$ alloy, have been chosen empirically. A systematic, fully first-principles study seems to be still missing.

\begin{figure}
\centering
\includegraphics[width=0.9\linewidth]{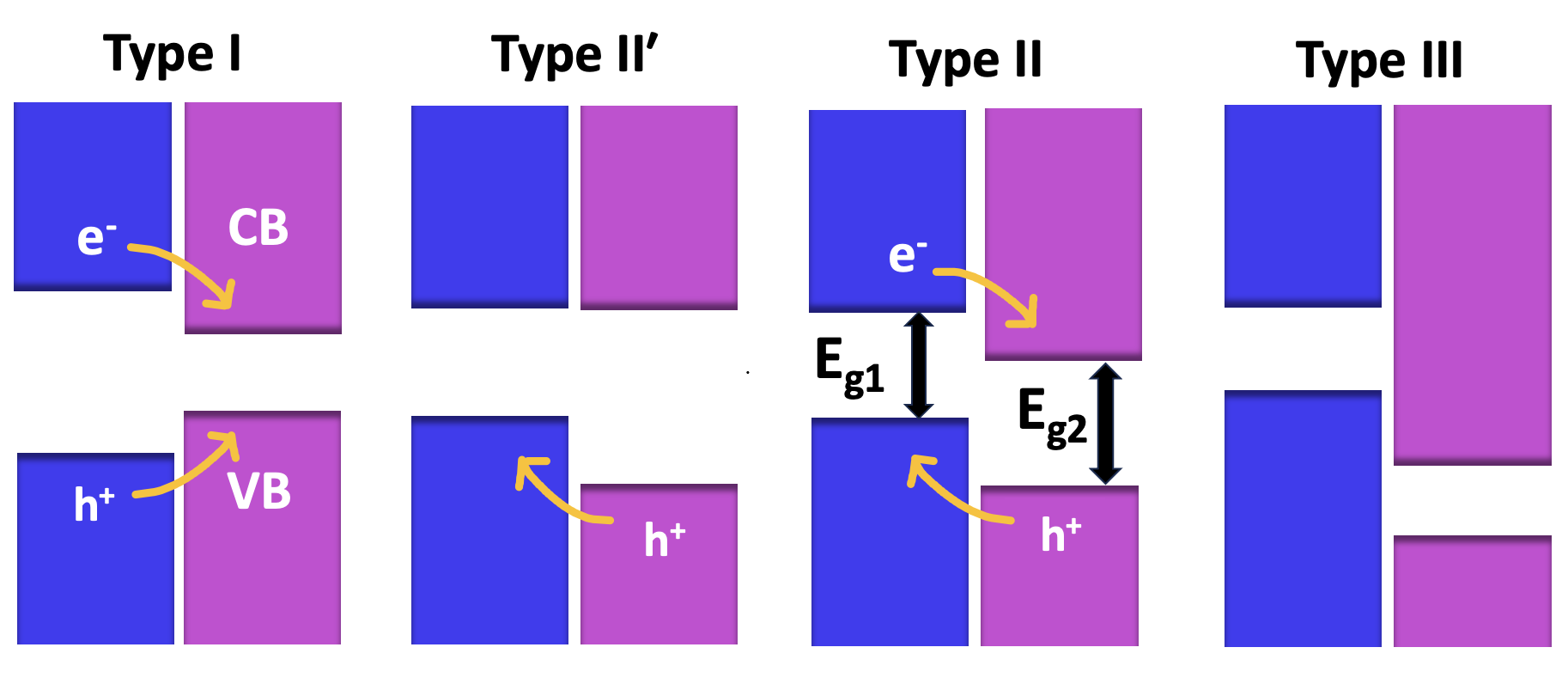}
\caption{Possible types of band alignment between two semiconductors M$_{1}$ and M$_2$ (indicated using blue and violet respectively). CB and VB indicate conduction and valence bands respectively and $E_{g1}$ and $E_{g2}$ are the band gaps of the two materials. The arrows indicate the possible directions of electron and hole flow on bringing the two semiconductors together. We have followed standard nomenclature for the types of band alignment, except for Type II$^{\prime}$ which is a new type of alignment we find between s-Si and Si$_{1-x}$Ge$_x$ for $x < 0.25$.}
\label{type_align}
\end{figure}

Motivated by this lacuna, in this paper, we present the results of Density Functional Theory (DFT) calculations of the properties of strained-Si/Si$_{1-x}$Ge$_{x}$ (denoted as s-Si/SiGe) heterostructures. 
In the literature, it is conventional to say that three different types of band alignment are possible on the basis of the relative positions of the CB and VB edges of the two materials M$_1$ and M$_2$ at the interface \cite{Kittel}. Based on the type of alignment  -- straddling, staggered, or broken -- these are labeled Type-I, Type-II, and Type-III respectively, and are shown schematically in Figure \ref{type_align}. In the figure, we have also introduced a fourth type of band alignment, which we call Type-II$^{\prime}$, for reasons that will become evident further below. 

In Figure \ref{type_align}, the arrows show how the type of band alignment determines the possible charge transfer across the interface. In Type-I band alignment, the sign of band offset in the CBs and the VBs is opposite. As a result, both electrons and holes can be transferred from M$_1$ to M$_2$. In contrast, in Type-II alignment, both CB and VB edges are higher in M$_1$ than the corresponding edges in M$_2$ so electrons are transferred from M$_1$ to M$_2$ and/or holes are transferred from M$_2$ to M$_1$. This can lead to the formation of a 2DEG in M$_2$ or a 2DHG in M$_1$. In Type-III, the VB edge in M1 is higher than the CB edge in M2, so no charge transfer can occur. Type-II$^{\prime}$ is a special case of Type-II where there is a negligible CB offset. Thus, only holes can be transferred from M$_2$ to M$_1$, leading to the formation of a 2DHG; a 2DEG cannot be formed.

It is difficult to experimentally determine the type of band alignment at a buried interface. One therefore has to turn to first-principles calculations. The determination of band alignments at semiconductor interfaces from first-principles is also challenging. The principal reason for this is the well-known `band gap problem' of DFT, whereby standard, computationally affordable DFT calculations severely underestimate band gaps and give unreliable results for band offsets and band alignments \cite{BA-methods}. Indeed, the popular Generalized Gradient Approximation (GGA) even predicts that bulk Ge should be a metal \cite{GGA-Eg-Ge}. Band gaps and band alignments can be computed more accurately by using hybrid functionals \cite{hse} or using the GW method \cite{Aryasetiawan_1998}; this however comes at a heavy computational cost. Moreover, calculations on alloy semiconductors require large unit cells, as well as averaging over a large number of structural configurations, thereby making the calculations even more expensive. A third problem is that calculating the band alignment at the interface involves either a comparison of band edge energies of bulk materials, which is tricky due to the lack of a common reference energy \cite{baldereschi} or directly considering the large heterostructure unit cell, thereby further increasing the computational cost. For these reasons, there have been very few previous theoretical investigations on s-Si/SiGe interfaces, and these employed either semi-empirical methods or standard DFT functionals with scissors corrections so as to match experimental band gaps. After this work was completed we became aware of a recent preprint where DFT has been used to estimate band alignment at s-Si/SiGe \cite{vegh2026_archiv}. The authors of this preprint used a  combination of hybrid functionals for the bulk materials with GGA calculations for the heterostructure to obtain band alignment; we will show below that our results differ from theirs in important ways.

An early Local Density Approximation (LDA) calculation by Van De Walle and Martin \cite{vande_martin} computed band alignments at s-Si/Ge, s-Ge/Si, s-SiGe/Si and s-SiGe/Ge interfaces. However, they did not consider the s-Si/SiGe interface which is of interest here. One previous study on this system was by Reiger and Vogl \cite{Reiger-Vogl} who performed empirical pseudopotential calculations on s-Si/Si$_{1-x}$Ge$_x$ for $x=0.5$, 0.85 and 1.0. Guided by these results, they predicted a Type-II band alignment for all values of $x$,  as did Vegh et al.~in their recent report \cite{vegh2026_archiv}. All other studies on SiGe-based interfaces appear to have looked at other systems, e.g., s-SiGe/Si and s-SiGe/Ge \cite{kurdi, people-bean,guldner}.
We note that if the formation of a 2DEG or 2DHG at an interface is to be directly demonstrated and investigated theoretically, it is crucial that the heterostructure be treated at a higher level of theory than standard DFT+GGA;  we are unaware of any such previous studies on s-Si/SiGe heterostructures. The only direct first-principles demonstrations of the formation of a 2DEG at a semiconductor-semiconductor interface have been for the polar s-AlN(or AlGaN)/GaN interface, where however, the origin of the 2DEG is different from that in our system \cite{GaN_AlN,GaN_AlGaN}.

Accordingly, here we present a fully first-principles study of s-Si/Si$_{1-x}$Ge$_x$ interfaces, using a hybrid functional. For the entire range of $x$, we compute the band gaps, band offsets and type of band alignment. 
While some of these results agree with those of previous authors who used more empirical and/or approximate methods, other results differ in significant ways. 
Further, for two carefully chosen values of $x$, we calculate the electronic structure, including the properties of the 2DEG or 2DHG formed in the presence of appropriate doping. 

\section{Results}\label{results}
\subsection{Band gaps of Si$_{1-x}$Ge$_{x}$ alloys and s-Si}
\begin{figure}[!h]
\centering
\includegraphics[width=0.75\columnwidth]{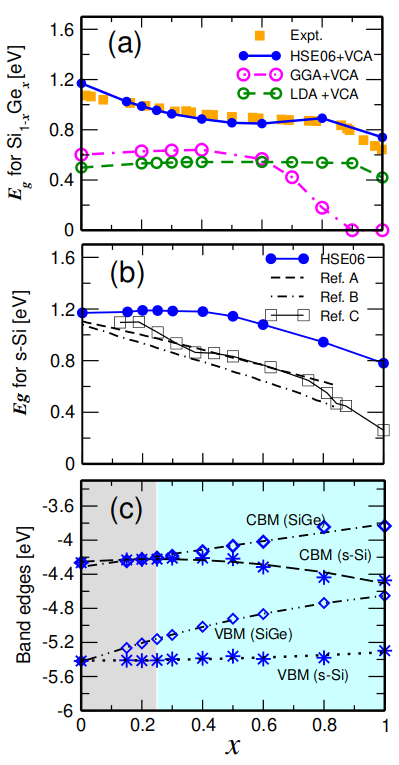}
\caption{Effect of variation of $x$ on (a) band gaps of Si$_{1-x}$Ge$_{x}$ (b) Si strained biaxially to the lattice constant of Si$_{1-x}$Ge$_{x}$ and (c) positions of band edges of s-Si and Si$_{1-x}$Ge$_{x}$. In (a), we compare our DFT results using the VCA and three different functionals, with those of experiment, taken from Ref.~\citenum{braunstein}. In (b), we compare our results to those of previous calculations reported in Refs.~A \cite{Yang}, B \cite{Reiger-Vogl}, C \cite{vegh2026_archiv}. In (c) VBM and CBM indicate the valence and conduction band edges, respectively, referenced to the vacuum. Gray and light blue areas indicate ranges of $x$ for which the band alignment is of Type II$^{\prime}$ and Type II, respectively. Blue, green and magenta symbols indicate DFT results obtained with HSE06, LDA and GGA, respectively.}
\label{fig:results_vs_x}
\end{figure}

In Figure \ref{fig:results_vs_x}(a) we show our DFT results (see Methods section for details) for how the band gap of the bulk Si$_{1-x}$Ge$_x$ alloy -- calculated using HSE06 and Virtual Crystal Approximation (VCA) -- varies with Ge concentration $x$. From this figure, we see that we obtain excellent agreement with experimental values \cite{braunstein}; this testifies to the validity of using the VCA for treating these systems. For comparison, we also show results obtained using either the local density approximation (LDA) or the generalized gradient approximation (GGA) instead of HSE06; we see that these functionals severely underestimate the band gaps (as expected), underlining the importance of using a hybrid functional to treat exchange-correlation. 

In Figure \ref{fig:results_vs_x}(b), we show how the band gap of biaxially strained-Si changes with $x$. We note that these results for s-Si make no use of the VCA. In the calculation, the in-plane (001) lattice constant of bulk Si was strained to that of Si$_{1-x}$Ge$_x$ at a given $x$, and the out-of-plane lattice constant $c$ was allowed to relax. Band gaps were then calculated for s-Si using the  HSE06 functional. The band gaps of s-Si show a negligible change at low values of $x$.  For higher values of $x$, the band gap decreases due to a bowing down of the conduction band edge [see Figure \ref{fig:results_vs_x}(c)].
We see that our results differ significantly from those of previous authors \cite{Yang,Reiger-Vogl,vegh2026_archiv}. We note that while the results for the band gap of s-Si are not directly presented in Ref.~\citenum{vegh2026_archiv}, we have deduced these values from their results for band gaps of SiGe and band offsets.
\subsection{Band alignment at s-Si/SiGe interface}

We now wish to determine the relative alignment of the band edges of Si$_{1-x}$Ge$_x$ and s-Si. The values of band energies as output from the DFT calculations for the two types of bulk systems cannot be directly compared, since the energies are referenced with respect to different zeros. To handle this problem, we follow a standard procedure \cite{tsai_ba,moses}, where additional slab calculations are performed, allowing the use of the vacuum energy level as a common reference. We note that this procedure implicitly assumes that the primary interface effects arise from the strain in s-Si, and eliminates the need for computationally expensive heterostructure calculations for each $x$. This procedure has been previously found to work well in predicting the type of band alignment at other semiconductor interfaces \cite{moses, tsai_ba, chang_ba, TanTansu2015}.

In Figure \ref{fig:results_vs_x}(c) we show our results for how the vacuum-aligned CB and VB edges of Si$_{1-x}$Ge$_x$ and s-Si shift with Ge concentration $x$. For $x < 0.25$, we see that though the VB edge of SiGe lies above that of s-Si, there is no offset between the CB edges. This has important implications for the physics at the interface as discussed below, which is why we introduce the nomenclature of Type II$^\prime$ for this kind of band alignment. We are not aware of any previous work that has reported such a band alignment at low values of $x$ for this system.  Our results imply that for $x < 0.25$, doping with an electron acceptor and/or appropriate electrostatic gating can result in the formation of a 2DHG at the interface; however, it is {\it not} possible to get a 2DEG at these low values of $x$ (see schematic diagrams in Figure S1 in the Supporting Information [SI] for an explanation.) 

In Section S2 of the SI, we compare these results with the recent results of Vegh et al. \cite{vegh2026_archiv}. We see that there are important quantitative and qualitative differences between the two sets of results. We also summarize the differences between our approach and theirs, which can explain this discrepancy. Their results show exactly the opposite at low values of $x$ -- viz., negligible offset in the VB edges but significant offset in the CB edges. Consequently, their result would imply that at low $x$, it should be possible to form a 2DEG but {\it not} a 2DHG, which is exactly the opposite of what we predict. We note that there is an experimental report of the formation of a 2DHG at  $x = 0.2$, with acceptor doping and/or gating \cite{whall,vogl_}, but no reports of the formation of a 2DEG at these low values of $x$. This gives us additional confidence in our approach and results. In Section \ref{sub:2dhg}, we will choose $x=0.1$ and an acceptor-type boron dopant as an example to directly demonstrate the formation of the 2DHG.

In contrast to the situation at low $x$, our results for band alignment for $x > 0.25$ imply that one can get either a 2DHG (centered toward the SiGe side of the interface) upon doping with an acceptor and/or electrostatic gating, or a 2DEG (centered toward the Si side of the interface) upon doping with a donor and/or appropriate gating (as explained in Figure S2 in the SI). Several experimental studies have reported that  a 2DEG is formed in donor-doped systems with  $x > 0.25$ \cite{2deg_30_sb_doped,2DEG_0.35,eff_mass_song,fidorra_2deg,eff_mass_measurement}. We point out that though a 2DHG can also be formed by acceptor-doping such systems, this 2DHG has the possible disadvantage of being primarily present on the disordered SiGe side of the interface. Further, very large values of $x$ are undesirable, due to the likelihood of structural defects arising from the large value of strain. Therefore, in  Section \ref{sub:2deg} we will choose $x=0.3$ and a donor-type phosphorus dopant to directly demonstrate the formation of the 2DEG.

\begin{figure*}[t]
   \includegraphics[width=0.8\linewidth]{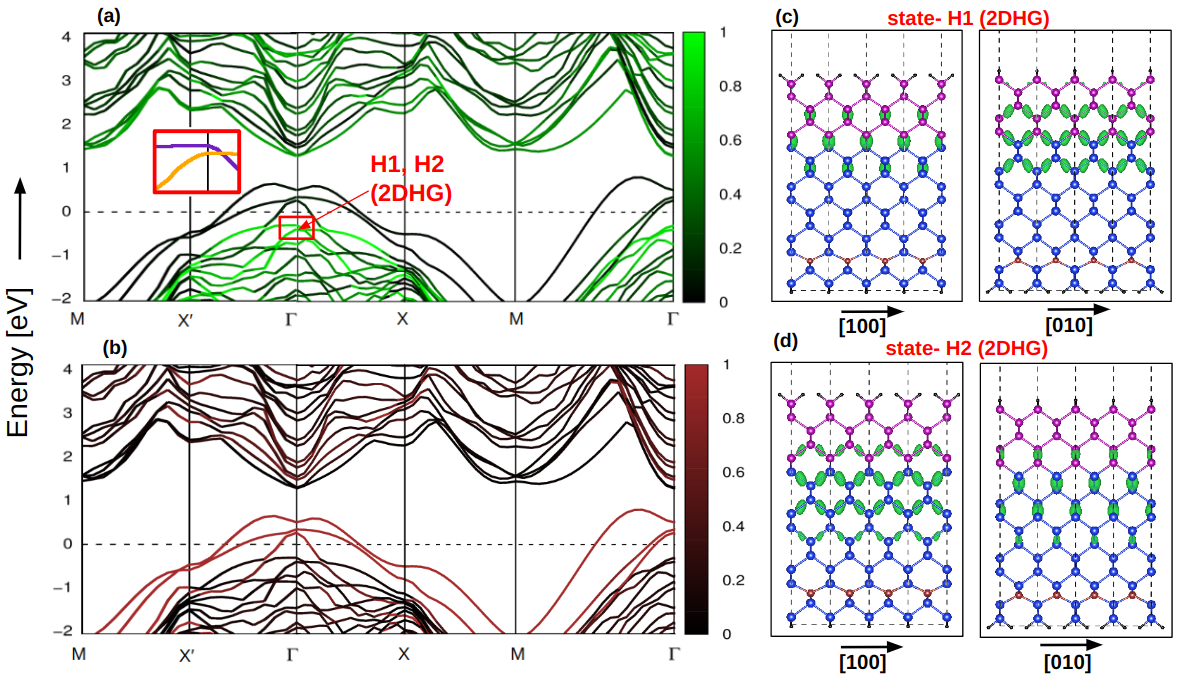}
   \caption{HSE06 band structure and charge densities of 2DHG states for B-doped s-Si/Si$_{0.9}$Ge$_{0.1}$.(a) Band structure with states projected onto atomic orbitals of SiGe atoms in the three layers adjacent to the interface; the color indicates the projection weight as per the scale on the right. The Fermi level is indicated by dotted horizontal lines. The hole-like states marked H1, H2 near $\Gamma$ arise dominantly from interfacial SiGe and correspond to 2DHG.
  (b) Same as (a), but with projections onto the B dopant atoms and the two SiGe layers in their immediate vicinity. We see three dopant bands that cross the Fermi level. (c),(d) Charge density distributions shown in green color (isosurface value = $9\times10^{-5}$) corresponding to the two nearly degenerate 2DHG states, shown for two different structural orientations (as indicated). s-Si atoms are shown in magenta, SiGe in blue, B dopant atoms in brown, and  H-atoms are tiny black spheres at the surfaces.}
   \label{fig:9010_2dhg}
\end{figure*}

\subsection{Heterostructures: Surface Slabs}\label{sub:hs_slabs}

We now verify the predictions made above, by performing hybrid DFT calculations on the following heterostructures: (i) s-Si/Si$_{0.9}$Ge$_{0.1}$ doped with  boron, (ii) s-Si/Si$_{0.7}$Ge$_{0.3}$ doped with phosphorus. We again emphasize that the use of the hybrid HSE06 functional is crucial here, to ensure the correct band gaps and band alignments at the interface -- in the absence of this, the properties of the 2DEG/2DHG are likely to be incorrect. 
 
The unit cells used for calculations on the heterostructures for Case (i) and Case (ii) can be seen in Figures ~\ref{fig:9010_2dhg} and \ref{fig:7030_2deg}, respectively. For computational reasons, we used slabs with $(1 \times 1)$ surface unit cells. For Case (i), the heterostructure unit cell consists of a substrate consisting of 12 layers of VCA Si$_{0.9}$Ge$_{0.1}$ stacked along the [001] direction. Above this substrate, we have five layers of s-Si. One of the substrate layers is substituted with dopant B atoms. For Case (ii),  we have a substrate consisting of eight layers  of VCA Si$_{0.7}$Ge$_{0.3}$, below nine layers of s-Si. One of the substrate layers is replaced with P dopant atoms.  A vacuum of 15 \AA\  was taken along the $c$-axis (normal to the surfaces of the slab). In both cases, both the upper and lower surfaces were passivated with hydrogen atoms to avoid dangling bonds. 

Figure S5 in the SI shows the two-dimensional interface Brillouin zone for these diamond-structure (001) slabs. Note that the two zone-edges X and X$^\prime$  are not equivalent for this structure.

\subsection{Case (i): Formation of 2DHG}\label{sub:2dhg}

In Figure \ref{fig:9010_2dhg} we present our main results for the electronic structure of B-doped s-Si$_{0.9}$Ge$_{0.1}$. Panels (a) and (b) display the calculated band structure, with projections shown onto near-interface SiGe orbitals and onto near-dopant orbitals, respectively. In this figure, the only bands that cross the Fermi level (dashed horizontal line) are dopant bands. However, by tuning the dopant concentration and/or appropriate electrostatic gating, the Fermi level can be tuned so that it shifts to slightly lower energies, so that the 2DHG bands (corresponding to the nearly-degenerate states `H1' and `H2' at the zone-center $\Gamma$) -- but no bulk-like bands -- cross the Fermi level. We have confirmed that the H-passivation has eliminated surface states. 

Panels (c) and (d) show the calculated charge density for the 2DHG states H1 and H2, respectively, viewed along both the $[100]$ and $[010]$ directions. We can see clearly that they are two-dimensional states. H2 has significant charge density primarily on the first few SiGe atomic layers near the interface, whereas H1 extends also into the first few s-Si layers near the interface. The two states are similar in nature, but H1 is more extended along the [010] direction and more localized along the [100] direction, while the reverse is true for H2.

The inset in Figure \ref{fig:9010_2dhg}(a) shows a zoomed-in picture of the band structure, focusing on the 2DHG states in the vicinity of the zone center ($\Gamma$). We use this to compute the effective masses $m^*$ for H1 and H2; these results are given in Table \ref{tab:eff_mass}. We see that $m^*$ is markedly anisotropic along the [010] and [100] directions, reflecting the extended or localized nature of the states [see Figures \ref{fig:9010_2dhg}(c) and \ref{fig:9010_2dhg}(d)]. The value of $m^*$ is compared to the experimentally available data for $x=0.13$ \cite{whall}; the agreement with the lower values of $m^*$ is excellent, as would be expected if both H1 and H2 are occupied by holes, when the anisotropy in the effective mass will not be apparent.  The flat band and very high effective masses of H1 along [100] and H2 along [010] indicate that holes in these states are strongly correlated and this may therefore lead to interesting physics. We also point out that if the carrier concentration can be tuned very precisely so that the Fermi level cuts H2 but not H1 near $\Gamma$, one may be able to achieve a 2DHG with an anisotropy so strong that it behaves essentially like a one-dimensional hole gas (1DHG). This can be further facilitated by the application of uniaxial strain. This lowers the symmetry between the [100] and [010] directions and will therefore increase the energy difference between H1 and H2. Previous authors have discussed the possibility of transforming a 2D carrier gas into 1D by the application of strain \cite{one_d,one_d_confine}. However, the underlying mechanisms described by them are different and do not exploit the strong in-plane anisotropy in effective masses of the carrier gas. 

\begin{table}[t]
\centering
\caption{Effective masses of 2DHG and 2DEG states in Case (i) B-doped s-Si/Si$_{0.9}$Ge${_{0.1}}$ and Case (ii) P-doped s-Si/Si$_{0.7}$Ge$_{0.3}$ systems, respectively, calculated at $\Gamma$ along $[100]\ (x)$ and $[010]\ (y)$ directions. Values are given in units of the free electron mass $m_e$. We note that the experimental value of $m^*$ for 2DHG is for $x=0.13$.}
\label{tab:eff_mass}
\setlength{\tabcolsep}{4pt}
\begin{tabular}{c c c c c }
\toprule
System & State & $m^*_{x} (\Gamma)$& $m^*_{y} (\Gamma$) & $m^*$(Expt.) \\
\midrule
Case (i): 2DHG
 & H1 & -9.1 & -0.22 & -0.23\cite{whall} \\
& H2 & -0.22 & -8.3 & \\
\midrule
Case (ii): 2DEG
& E1 & 0.19 & 0.20 & 0.20\cite{eff_mass_song}  \\
& E2 & 0.22 & 0.24 & \\
\bottomrule
\end{tabular}
\end{table}

\subsection{Case (ii): Formation of 2DEG }\label{sub:2deg}
\begin{figure*}[t]
\includegraphics[width=0.8\linewidth]{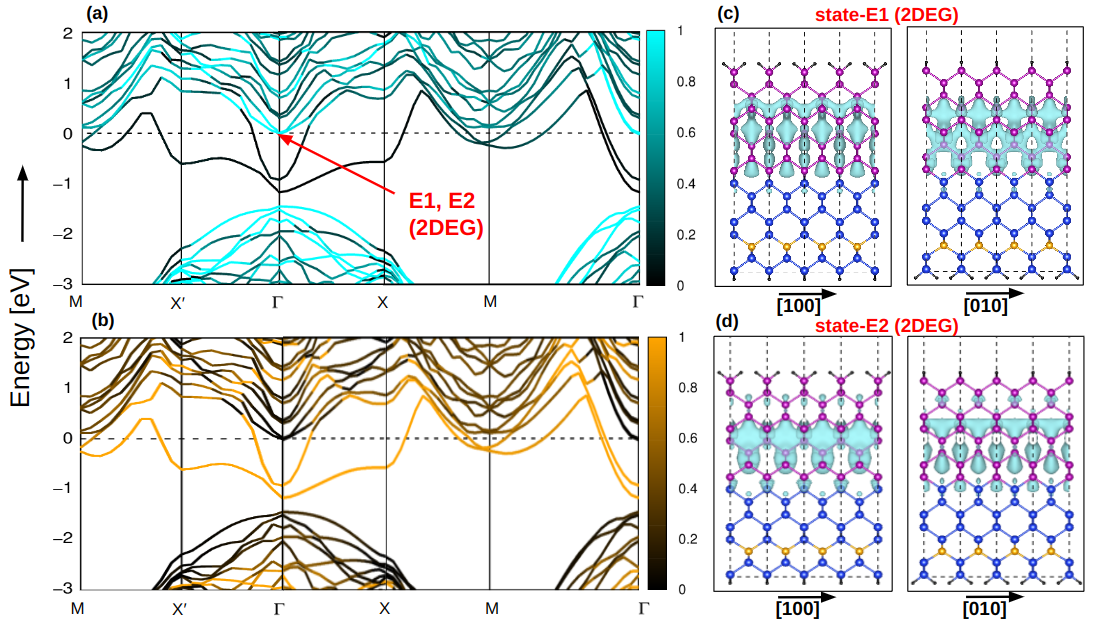}
\caption{HSE06 band structure and charge densities of 2DEG states for P-doped s-Si/Si$_{0.7}$Ge$_{0.3}$. (a) Band structure with states projected onto atomic orbitals of s-Si atoms in the three layers adjacent to the interface; the color scale indicates the amount of projection. The Fermi level is indicated by dotted horizontal lines. The states marked E1 and E2 near $\Gamma$ arise primarily from interfacial s-Si states and correspond to 2DEG states. (b) Same as (a), but with projections onto the P dopant atoms and the two SiGe layers in their immediate vicinity. We see two dopant bands that cross the Fermi level. (c),(d) Charge density distributions shown in cyan color (isosurface value = $2\times10^{-4}$) corresponding to the two nearly degenerate 2DEG states labeled E1 and E2 at $\Gamma$, shown for two different structural orientations (as indicated). s-Si atoms are shown in magenta, SiGe in blue, P dopant atoms in deep yellow, and  H-atoms are tiny black spheres at the surfaces.}
\label{fig:7030_2deg}
\end{figure*}
% %
In Figure \ref{fig:7030_2deg}, we present the electronic structure of P-doped s-Si/Si$_{0.7}$Ge$_{0.3}$. Panels (a) and (b) show the calculated band structure, with projections onto near-interface Si orbitals and near-dopant orbitals, respectively. The Fermi level ($E_{F}$)  is crossed by only dopant-derived bands and just touches the 2DEG states E1 and E2. These states will become occupied on shifting $E_{F}$ slightly upward, e.g., by gating. Panels (c) and (d) display the charge density corresponding to E1 and E2, respectively, viewed along the $[100]$ and $[010]$ directions. As for the 2DHG in Case (i), these states are again clearly two-dimensional, with however the charge density now predominantly localized on the s-Si layers near the interface. While both E1 and E2 share similar characteristics, E1 is slightly more extended along the $[010]$ direction and more confined along $[100]$, whereas E2 exhibits the opposite anisotropy.

The effective masses $m^*$ for E1 and E2 are given in Table~\ref{tab:eff_mass}. We see that $m^*$ is slightly anisotropic along the [100] and [010] directions, with a value very close to that measured experimentally \cite{eff_mass_song}, we are not aware of any previous attempts to compute $m^*$ for this system.

\section{Summary and Conclusions}\label{conclusion}
In this paper, we have used DFT calculations, performed using the hybrid HSE06 functional and the VCA, to revisit the problem of engineering two-dimensional electron or hole gases at s-Si/SiGe interfaces.

Our calculated band gaps for bulk Si$_{1-x}$Ge$_x$ as a function of $x$ are in excellent agreement with experiment, supporting the validity of our approach.
For s-Si, our results for the change in band gap with strain differ -- both qualitatively and quantitatively -- from those of previous authors.

We find that at low Ge concentration $(x < 0.25)$, the band alignment at the s-Si/SiGe interface is such that there is no band offset in the conduction bands, but there is an offset in the valence bands, with the valence band edge for SiGe positioned higher than that for s-Si. We have named this type of band alignment Type II$^\prime$, as we feel it is important to distinguish it from Type II alignment, since it leads to different physics at the interface. In particular, it implies that with acceptor-doping and electrostatic gating, one can get a 2DHG at the interface. However, it is not possible to engineer a 2DEG at this interface. This finding does not appear to have been previously reported in the literature. However, it is consistent with earlier experimental reports \cite{whall}.

For higher Ge concentrations ($x > 0.25$), we find a Type II band alignment, in agreement with previous authors \cite{Reiger-Vogl,vegh2026_archiv}, although the band offsets calculated by us are quite different in magnitude. This implies that donor-doping + gating will result in the formation of a 2DEG on the s-Si side of the interface, whereas acceptor-doping + gating will result in a 2DHG on the SiGe side. We note that in the experimental literature, it is common to use values around $x = 0.3$ to obtain a 2DEG \cite{2DEG_0.35,2deg_30_sb_doped,eff_mass_song} -- our results show why this value works well, since $x$ is then high enough to ensure Type II band alignment at the interface, yet low enough that strain-induced structural defects are less likely to be present.

We next go on to directly demonstrate the validity of the above predictions, by performing calculations on appropriately doped s-Si/SiGe heterostructure slabs, with the value of $x$ chosen suitably ($x = 0.1$ and $x = 0.3$). We emphasize that these heterostructures are treated at HSE06 level; to the best of our knowledge this has not been previously done, presumably due to reasons of computational cost.

For B-doped s-Si/Si$_{0.9}$Ge$_{0.1}$ we have found the electronic band structure, and see that we obtain two 2DHG bands near the Fermi level. 
These 2DHG states are found to be strongly anisotropic, with markedly different effective masses along [100] and [010], and feature almost flat bands along one of these two directions. This raises the possibility of achieving a 1DHG, as well as possibly exotic physics driven by strong correlations. The experimentally measured effective mass is in excellent agreement with the two lower values obtained by us. Similarly, for P-doped s-Si/Si$_{0.7}$Ge$_{0.3}$, the electronic band structure now features instead two almost-degenerate 2DEG bands near the Fermi level. The values of effective mass for the 2DEG are only slightly anisotropic, and are in excellent agreement with experimentally reported values. 

We have plotted the charge densities for these 2DHG and 2DEG states, and confirmed their two-dimensional nature, and the confinement of carriers to the first few near-interface atomic layers.

This work constitutes a proof-of-concept, 
that it is now possible to use first-principles calculations to directly study the electronic properties of two-dimensional electron and hole gases at the interfaces of semiconductor heterostructures. Some of this information is difficult to access experimentally, and thus first-principles calculations are very necessary. Our work emphasizes that it is vital to treat exchange-correlation interactions at a high level so as to correctly calculate such properties. Possible extensions of this work would be to investigate other doping concentrations and profiles, and the efficacy of different dopants, as well as mimic the effects of electrostatic gating by the application of electric fields. (These extensions would, however, be more computationally expensive).  Finally, it would be of great interest to use results like those presented here as input to calculations of the transport properties of these 2D electron and hole gases. 

\section{Methods}\label{methods}
First principles Density Functional Theory (DFT) calculations were performed using Projector Augmented Wave (PAW) \cite{paw} pseudopotentials as implemented in the Vienna Ab-initio Simulation Package (VASP) \cite{vasp,vasp-2,vasp-3} with a plane wave cutoff of 380 eV. To accurately obtain the values of the band gaps and band alignments, we used the Heyd-Scuseria-Ernzerhof (HSE06) hybrid functional \cite{hse}; this has mixing parameter $\alpha=0.25$ and screening length $\omega=0.2$ \AA $^{-1}$. It was verified that these values of $\alpha$ and $\omega$ best reproduce the experimental band gaps of Si, Si$_{0.7}$Ge$_{0.3}$, and Ge.

Structural relaxations were performed at the Perdew-Burke-Ernzerhof \cite{gga}(PBE)-GGA level, and all subsequent electronic structure calculations were carried out using HSE06. All atomic coordinates were allowed to relax with a force convergence threshold of 0.001 eV/\AA. 
All calculations were performed using experimental lattice constants.
Spin-orbit interactions are not included, as they are expected to make only a minor correction to the band gaps and band offsets presented here \cite{soc_yu}.

\begin{table}[t]
\begin{center}
\caption{Band gaps $E_g$ of bulk Si and Ge calculated using HSE06 and GGA functionals, compared with experimental values. Experimental values (extrapolated to ${T}= 0\ \rm{K}$) are obtained from Refs.~\cite{madelung2004semiconductors, Kittel}.}
\begin{tabular}{c c c c}
\hline
System & $E_g \rm{(HSE06)}$ & $E_g \rm{(GGA)}$ & $E_g \rm{(Expt)}$\\  
 & [eV] & [eV] & [eV] \\ \hline
bulk Si & $1.17$ & $0.6$ & $1.17$ \\ 
bulk Ge & $0.74$ & $0$ & $0.744$ 
\\
\hline
\end{tabular}
\label{tab:band_gaps}
\end{center}

\end{table}  

The values of band gaps of bulk Si and bulk Ge calculated using HSE06 are shown in Table \ref{tab:band_gaps}. These values are in excellent agreement with the experimental values \cite{Kittel,madelung2004semiconductors}.

Since Si and Ge are chemically similar, the Virtual Crystal Approximation (VCA) can be used for Si$_{1-x}$Ge$_x$ alloys \cite{vca}. It is known previously that this is a good approximation for SiGe alloys \cite{gironcoli_vca, hahn_vca, ikoni_vca, langueur_vca, fiorentino_vca}. Treating alloys using the VCA considerably reduces unit cell sizes for both bulk and heterostructures, and obviates the need for averaging over structural configurations.

Band alignment was determined by referencing all energies to the vacuum level in a two-step process \cite{tsai_ba,moses}. First, the CB edge
and VB edge of each bulk material (s-Si and SiGe) were calculated relative to the bulk macroscopic average electrostatic potential \cite{baldereschi}. Next, 5-layer slab calculations were used to determine the difference between the bulk-like electrostatic potential and the vacuum level. These values remain unchanged with thicker slabs. Combining these results gives the band-edge positions relative to the vacuum level, enabling direct comparison of VB and CB offsets and identification of the type of band alignment.

\section{Acknowledgments}
We acknowledge funding support from the National Quantum Mission through the Foundation for Quantum Computing Innovation, an initiative of the Department of Science and Technology, Government of India. We acknowledge the support of the National Supercomputing
Mission (NSM) of India for the use of the Param Yukti Supercomputing cluster, JNCASR. The authors also acknowledge fruitful discusions with Madhu Thalakulam.

\section{Conflict of Interest}
The authors declare no conflicts of interest.

\section{Data Availability}
The data can be made available upon reasonable request.
\bibstyle{apsrev4-2}
\bibliography{bibliography}

\clearpage
\onecolumngrid

\section*{Supporting Information}
\setcounter{section}{0}
\renewcommand{\thesection}{S\arabic{section}}
\linespread{1.4}\selectfont
\begin{center}
\textbf{Properties of 2D Electron or Hole Gases at Tailored s-Si/SiGe Interfaces: A First-Principles Investigation}
\end{center}

\begin{center}
Garima Ahuja$^{1}$, Farha S$^{2,3}$, Tanuja Shridhar Joshi$^{1}$, Anil Shaji$^{2,3}$, \\Shobhana Narasimhan$^{*1,4}$

\end{center}
\begin{center}
$^{1}$ Theoretical Sciences Unit and School of Advanced Materials, Jawaharlal Nehru Centre for Advanced Scientific Research, Jakkur, Bengaluru 560064, India \\
$^{2}$ School of Physics, Indian Institute of Science Education and Research Thiruvananthapuram, Maruthamala PO, Vithura, Kerala, 695551, India \\
$^{3}$ Center for High Performance Computing, Indian Institute of Science Education and Research Thiruvananthapuram, Maruthamala PO, Vithura, Kerala, 695551, India \\
$^{4}$ International Centre for Theoretical Sciences, Shivakote, Hesaraghatta, Bengaluru 560089, India
\end{center}

\newpage
\renewcommand{\thefigure}{S\arabic{figure}}
\setcounter{figure}{0}
\section{Schematic diagrams showing band bending leading to 2DEG/2DHG}
Here we present schematic diagrams to show how band bending leads to the formation of a 2DEG or 2DHG in s-Si/Si$_{1-x}$Ge$_x$ heterostructures. In the Figs.~S3, S4, and S5, $E_C$ and $E_V$ denote the positions of conduction and valence band edges, respectively, for the two materials before they are joined together to form a heterojunction, and $E_C$$^{\prime}$ and $E_V$$^{\prime}$ denote the new positions of conduction and valence band edges after the materials are joined. For our system, blue color represents SiGe and purple represents s-Si. $\Delta CB$, $\Delta VB$, and $E_{g}$ denote the conduction band offset, valence band offset, and band gap respectively. $E_F$ denotes the Fermi level in each of the two materials.
\begin{figure}[h]
    \centering    \includegraphics[width=0.9\linewidth]{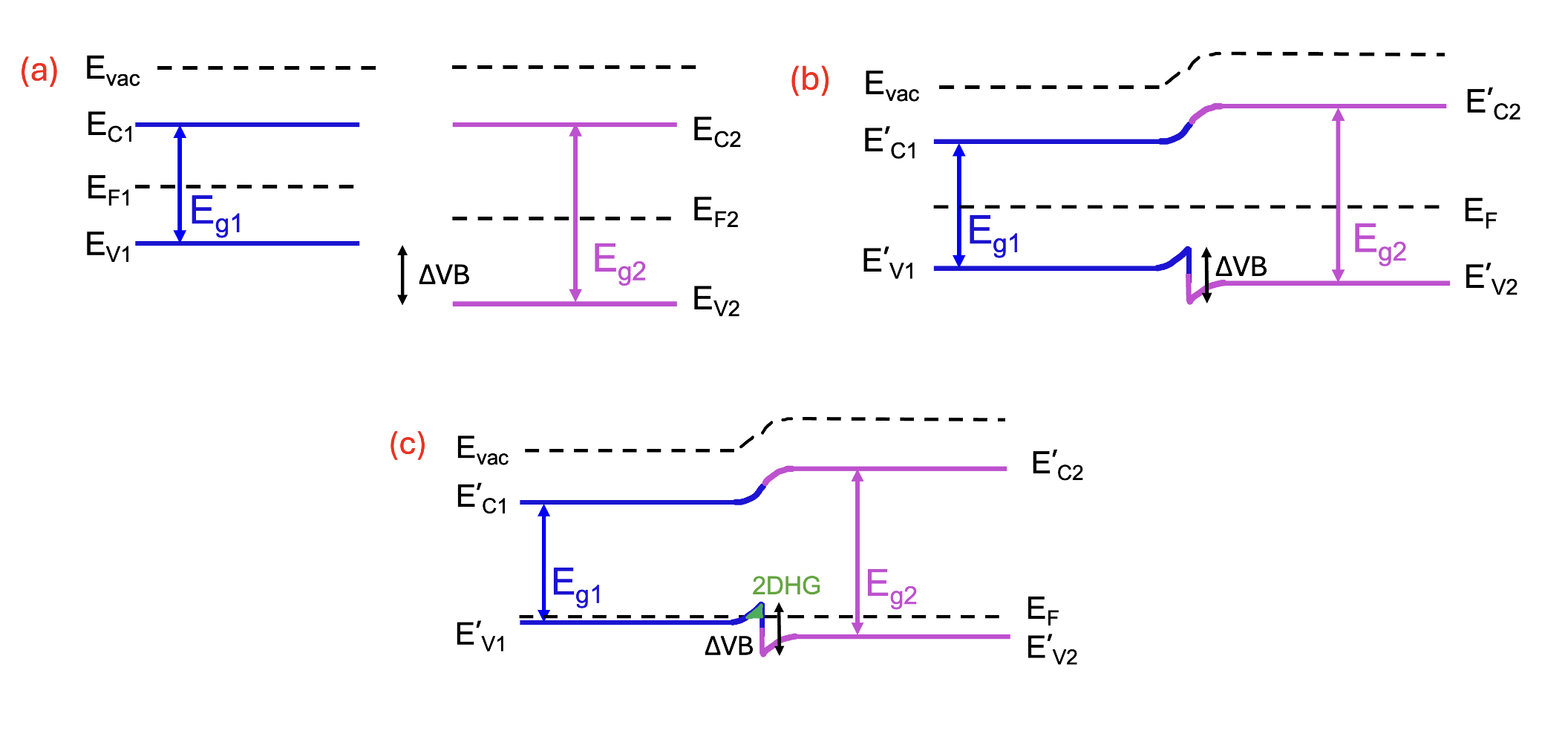}
    \caption{Schematic diagrams showing band bending and formation of a 2DHG for Type-II$^{\prime}$ band alignment (situation when $x<0.25$). (a) Conduction band edges, valence band edges, and Fermi levels of the two materials before contact. There is a valence band offset but no conduction band offset, (b) alignment of Fermi levels after contact, leading to band bending near the interface, (c) shift of Fermi level towards valence band on p-type doping in one of the materials, leading to formation of a 2DHG.}
    \label{fig:case1}
\end{figure}

Figure \ref{fig:case1} depicts the case when $x < 0.25$.
There is no conduction band offset between CB edges of s-Si and SiGe but there is a valence band offset. After the two materials are joined together, the Fermi levels of the two must align. This causes the bending of conduction and valence band edges, and also the vacuum levels. This is shown in Figure \ref{fig:case1}(b). Due to the presence of offsets in the valence bands, there is a formation of a kink in the valence band edge. However, due to the absence of a conduction band offset, no such kink is present in the conduction band edge at the interface.
There is therefore the possibility of obtaining a 2DHG, by accumulating holes within the kink upon moving the Fermi level downward by p-type doping. This is shown in Figure \ref{fig:case1}(c). On the other hand, there cannot be a 2DEG even if the Fermi level were to be moved toward the conduction band by n-type doping, because of the absence of a kink in the conduction band edge.

\begin{figure}
    \centering
    \includegraphics[width=0.9\linewidth]{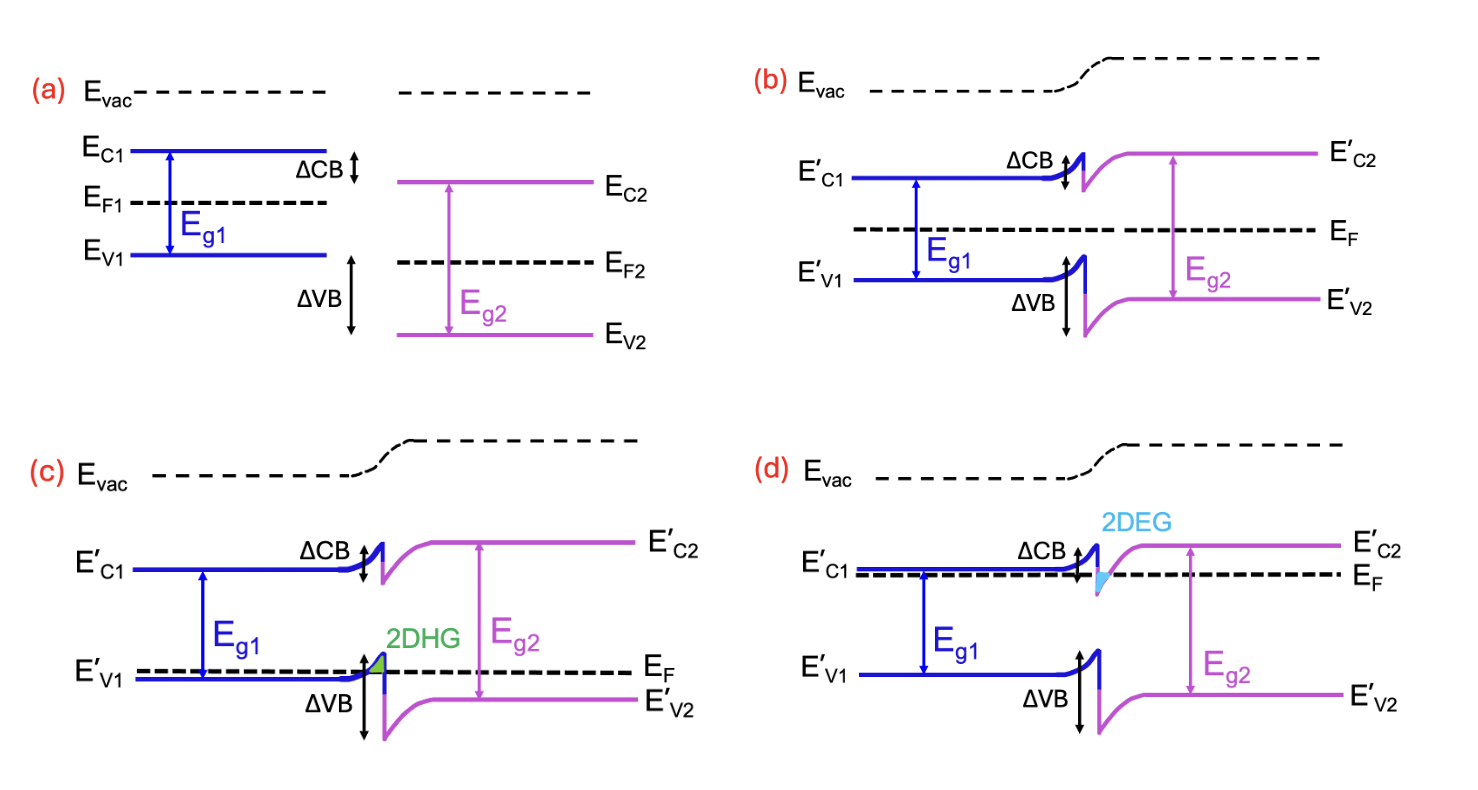}
    \caption{Schematic diagrams showing band bending and formation of a 2DHG or a 2DEG for Type-II band alignment (situation when $x>0.25$). (a) Conduction band edges, valence band edges, and Fermi levels of two materials before contact, (b) alignment of Fermi levels after contact, leading to band bending near the interface, (c) shift of Fermi level towards valence band on p-type doping in one of the materials, leading to formation of a 2DHG, (d) shift of Fermi level towards conduction band on n-type doping in one of the materials, leading to formation of a 2DEG.}
    \label{fig:case2}
\end{figure}

When $x > 0.25$ there are offsets in both the conduction and valence band edges, and therefore there is a kink at the interface in both the conduction and valence band edges. Thus, it is possible to form either a 2DEG or 2DHG by suitable doping to move the Fermi level and occupy the kink in the conduction band or valence band, respectively.
The two possibilities are shown in Figures \ref{fig:case2}(c) and (d). 

Figures \ref{fig:case1} and \ref{fig:case2} correspond to the Type-II$^\prime$ and Type-II band alignments found by us for $x < 0.25$ and $x > 0.25$ respectively. 

\begin{figure*}
    \centering    \includegraphics[width=0.9\linewidth]{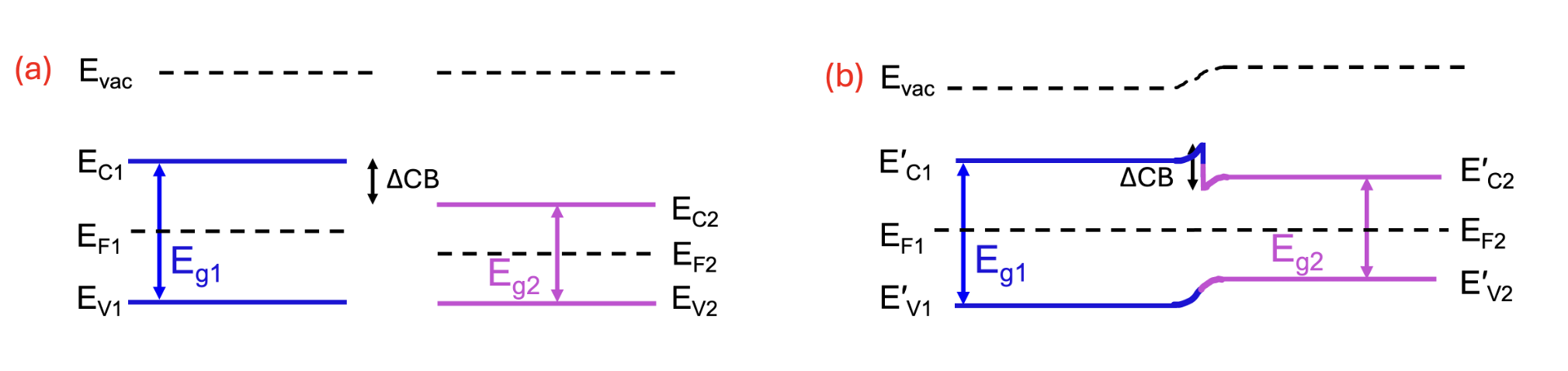}
    \caption{Schematic diagrams showing band bending for the case when there is no valence band offset, but a conduction band offset. (a) conduction band edges, valence band edges, and Fermi levels of two materials before contact, (b) alignment of Fermi levels after contact, leading to band bending near the interface. No kink is formed in valence band, thus a 2DHG cannot be obtained.}
    \label{fig:case3}
\end{figure*}

These predictions can be contrasted with the results of Vegh et al. \cite{vegh2026_archiv} who find no offset in the valence band but an offset in the conduction band for low $x$. This situation is depicted in Figure \ref{fig:case3}. In this case, we see that the kink is formed in the conduction band but not in the valence band at the interface. It is then possible to form a 2DEG by n-type doping but it is not possible to form a 2DHG even after p-type doping.  
\newpage
\section{Comparison of Band Alignment Results with Previous Work by Vegh et al. }

\begin{figure}[h!]
    \centering
    \includegraphics[width=0.5\linewidth]{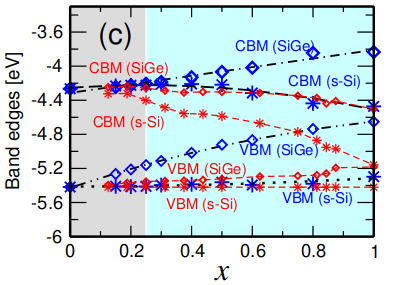}
    \caption{Positions of band edges of s-Si and Si$_{1-x}$Ge$_{x}$. CBM and VBM of SiGe (diamond symbols) and s-Si (star symbols) are indicated for our results in blue and those of Vegh et al.\cite{vegh2026_archiv} in red colors, respectively. According to our results, Gray and light blue areas indicate ranges of $x$ for which the band alignment is of Type II$^{\prime}$ and Type II, respectively.}
    \label{fig:ba_compare}
\end{figure}
As mentioned in the main text, our results for band alignment between s-Si and SiGe differ significantly from those reported in a recent preprint\cite{vegh2026_archiv}. In Figure \ref{fig:ba_compare} we compare our results for the positions of band alignments with theirs. The positions of the band edges, as reported in Ref.~\cite{vegh2026_archiv} have been extracted from their reported values of band offsets, and the band gaps of SiGe. On comparing our results (blue symbols) with theirs (red symbols), we find important qualitative as well as quantitative differences.

We see that their valence band offsets are small for the entire range of $x$, and negligible at lower values of $x$, whereas our results show that the valence band offsets vary significantly and increase with $x$. This difference may be due to the differences in procedure used. 

In contrast, for the conduction bands we find essentially zero offset at small $x$, whereas they find that the offset increases markedly over the entire range of $x$. 

As already mentioned in the main text, the results in this figure suggest that our results for the band gaps for SiGe are similar to theirs, their results for the band gap of s-Si are much smaller than ours, except for $x \sim 0$. 

Our approach differs from theirs in other ways too, which might contribute to the differences in Figure \ref{fig:ba_compare}. We have treated the SiGe alloys using the VCA, whereas they have used Special Quasirandom Structures (SQS), averaging over a small number of configurations. To align the band edges, we have used an all HSE06 bulk-slab method to enable the use of the vacuum energy as the common reference level. They have instead performed PBE calculations on a repeating superlattice, which is used to find the difference between the bulk electrostatic potentials of the two materials. They seem to calculate the valence band offsets at PBE level, whereas we use the HSE06 for this. We note that their use of the SQS results in appreciable variations in the value of the the bulk electrostatic potential in the alloy region. 

In consequence, we find a Type II$^\prime$ band alignment for $x < 0.25$ and Type II for $x > 0.25$, predicting that with doping and gating only a 2DHG can be formed for $x < 0.25$, whereas their results imply that only a 2DEG can be formed for low $x$, due to the lack of offset in the valence bands (see Figure \ref{fig:case3}). We note that there is a previous experimental report of the formation of a 2DHG at low $x$ \cite{whall}. But to the best of our knowledge, a 2DEG has not been reported for such low values of $x$. This gives us confidence in our results as presented in Figure \ref{fig:ba_compare}. Further, our predictions are directly verified by HSE06 calculations as demonstrated in the main text. 

For larger values of $x$, their predictions and ours are qualitatively similar but quantitatively different. 
\newpage

\section{Brillouin zone for diamond(001) interface}
\begin{figure}[!h]
    \centering
    \includegraphics[width=0.3\linewidth]{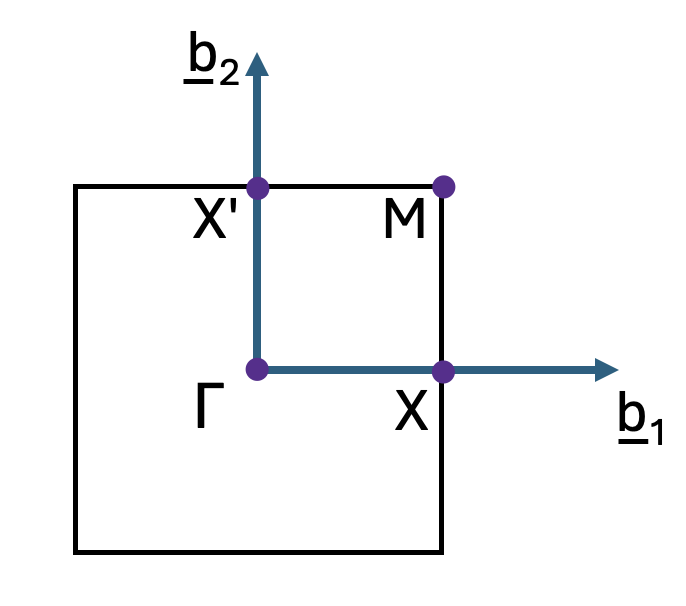}
    \caption{2D interface Brillouin zone for diamond(001) surface. \underline{b$_{1}$} and \underline{b$_{2}$} are the reciprocal lattice vectors. X, X$^\prime$, $\Gamma$, and M are the high symmetry points.}
    \label{fig:bz}
\end{figure}

The Figure \ref{fig:bz} shows the two-dimensional interface Brillouin zone for the diamond-structure (001) slabs used in the calculations. The high symmetry points X and X$^\prime$ are not equivalent for the structure. 

\newpage
\textbf{References}

[S1] N. M. Vegh, P. Philippopoulos, R. J. Prentki, W. Zhang, Y. Zhu, F. Beaudoin, and H. Guo, First-principles predictions of band alignment in strained Si/Si$_{1-x}$Ge$_x$ and Ge/Si$_{1-x}$Ge$_x$ heterostructures. arXiv:2603.13219 (2026).

[S2] T. E. Whall, A. D. Plews, N. L. Mattey, P. J. Phillips, U. Ekenberg, Effective mass and band nonparabolicity in remote doped Si/Si$_{0.8}$Ge$_{0.2}$ quantum wells. Applied Physics Letters, 66 (20): 2724–2726 (1995).
\end{document}